# Dynamic spectrum sharing game by lease[1]

Cuilian Li[1,2], Zhen Yang[1], Jun Li[2], Feng Tian[1*]

1. Institute of Signal Processing and Transmission at Nanjing University of Posts and Telecommunications, P.O. Box 214, 66 Xin Mofan Rd., Nanjing, Jiangsu 210003, China
2. Zhejiang Wanli University, Ningbo, Zhejiang 315100, China

**Abstract**

We propose and analyze a dynamic implementation of the property-rights model of cognitive radio. A primary link has the possibility to lease the owned spectrum to a MAC network of secondary nodes, in exchange for cooperation in the form of distributed space-time coding (DSTC). The cooperation and competition between the primary and secondary network are cast in the framework of sequential game. On one hand, the primary link attempts to maximize its quality of service in terms of signal-to-interference-plus-noise ratio (SINR); on the other hand, nodes in the secondary network compete for transmission within the leased time-slot following a power control mechanism.

We consider both a baseline model with complete information and a more practical version with incomplete information, using the backward induction approach for the former and providing approximate algorithm for the latter. Analysis and numerical results show that our models and algorithms provide a promising framework for fair and effective spectrum sharing, both between primary and secondary networks and among secondary nodes.

## 1. Introduction

The concept of dynamic spectrum access is currently under investigation as a promising paradigm to achieve efficient use of the frequency resource. Different sharing models allow the coexistence of licensed (primary) users and unlicensed (secondary or cognitive radios, CR) in the same bandwidth through different schemes. The common models [1,2] do not assume any interaction between primary links and CRs, while the property-rights(or spectrum leasing) [1,3] models allow the primary terminals owning a given bandwidth lease it for a fraction of time to CRs in exchange for appropriate remuneration. Although both kinds of models have limitations, the property-rights models have seldom been analyzed in the communication engineering literature. In the effort to merge techniques and strategies, we propose a dynamic implementation of the property-rights model, and thus improve spectrum efficiency.

The efficiency of an implementation of the property-rights model depends on concrete techniques such as resource allocation and management, cooperative communication, spectrum access and pricing schemes etc. R. D. Yates provides a framework of uplink power control for the cellular wireless communication systems in [4], to minimize the transmitting

---

[1*] Cuilian Li. e-mail: licl@njupt.edu.cn.

power while guarantee the target SINRs. Authors of [5,6] put forward opportunistic power control frameworks which are applicable to systems supporting opportunistic communications and yield significant improvement in throughput when compared with conventional target tracking approach in [4]. Authors of [7,8] define cost or utility functions that consist of weighted sum of power and achievable QoS, gain substantial power savings while slightly reduce the achieved SINR, compared to the power balancing algorithm. Authors of [9-12] aim at improving convergence rate and provide optimizing technologies such as combination-optimization, successive over-relaxation method and exponential type of power update function etc.[13,14] survey various coding schemes and protocols used in cooperating communication.[15-19] model particular distributive resource allocation problems as potential games or S-modular games where there exist coupling among the strategy vectors of all players and analyze the convergence of their algorithms. B. Wang etc. [20, 21] propose a Stackelberg game theoretic framework for distributed resource allocation over multiuser cooperative communication networks to improve the system performance. [22, 23] focus on fairness and efficiency of their resource allocation algorithms in scenes where nodes have various QoS requirements. Although results in the above papers can increase spectrum efficiency to some extent, they focus on either unlicensed spectrum or underlay or overlay [24] mode, neither of them considers interaction between primary links and CRs.

Authors of [1, 3] compare property-rights with common models, O. lleri etc. [2] put forward two property-rights model of dynamic spectrum access, yet limited to illustrative explanation. We focus on combining techniques with spectrum leasing model and propose a spectrum sharing framework. By helping the primary link through cooperating communication and thus improving its QoS, secondary terminals are granted the use of a given spectral resource by the incumbent primary. We investigate both scenes with or without complete information. Our works differ from the work of O. Simeone etc.[25] in three aspects: firstly, our secondary network is a MAC system while O. Simeone's an ad hoc; secondly, O. Simeone etc. choose decode-and-forward schemes (DF) and assert that amplify-and-forward ones (AF) are not applicable in their framework, we instead adopt AF for the sake of comparison despite both AF and DF schemes fit our framework; and thirdly, we provide feasible approximate solution for our models while O. Simeone etc. only give the solvable conditions of their model.

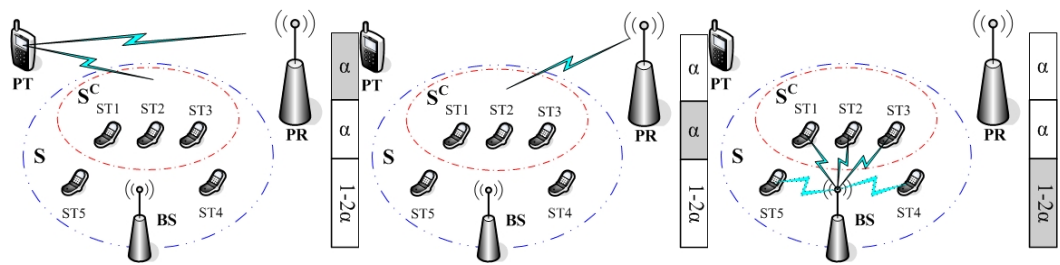

*Fig. 1*. Secondary spectrum access through cooperation-based dynamic spectrum leasing

*(a)* primary transmission  *(b)* DSTC cooperation  *(c)* secondary transmission

## 2. System model

Overall, the considered framework is characterized by a hierarchical structure, where three entities (primary link between primary transmitter (PT) and receiver (PR), secondary base-station (BS) and CRs take turn to act. The high level is modeled as a sequential game between primary and cognitive network. On one hand, one player (the primary link) optimizes its

———

strategy (leased time) based on the knowledge of the effects of its decision on the behavior of a second player (the secondary network, including its BS and CRs); on the other hand, cognitive BS decides whether to cooperate and if so, the cooperative scale (number of cooperating relays to be allowed) based on the leased time while cognitive CRs decide the transmitting power. The low level involves CRs competing among themselves for transmission within the leased time-slot, following a distributed opportunistic power control mechanism.

*2.1. Medium access control layer*

We consider the system sketched in Fig. 1, where a PT communicates with the intended PR in slots of NS symbols. In the same bandwidth, a secondary multiple access network composed of $K=|\mathbf{S}|$ CRs is seeking to exploit possible transmission opportunities. The primary link grants the use of the bandwidth to CRs in exchange for cooperation so as to improve the quality of the communication link to its receiver PR. Although metrics such as link capacity, bit-error rate (BER) or outage probability etc. may all be qualified candidates, they are all monotone functions of SINR. We choose SINR as utility metric, both for its basic, and simplicity.

In particular, the primary decides whether to use the entire slot for direct transmission to PR or to employ cooperation. In the latter case, a fraction α of the slot (0<α<1/2) is used for transmission from PT to PR and to CRs in $\mathbf{S}^C \subset \mathbf{S}$ (see Fig. 1-(a)). The second fraction α of the slot is dedicated to cooperation: The set $\mathbf{S}^C$ of cognitive relays form a distributed $|\mathbf{S}^C|$ antenna array that cooperatively relay the primary codeword through DSTC [13,14] towards PR (Fig. 1-(b)). And in the last subslot of duration (1-2α) NS symbols, all the active CRs are allowed to transmit their own data (Fig. 1-(c)), in which the transmission scheme amounts to multiple access channels with distributed power control [4-6].

*2.2. Physical layer*

The channels between nodes are modeled as independent proper complex Gaussian random variables, invariant within each slot, but generally varying over the slots (i.e., Rayleigh block-fading channels). We use the following notation to denote the instantaneous fading channels in each block: $h_P$ denotes the complex channel gain between PT and PR; $h_{PS,i}$ the channel gain between PT and secondary transmitter STi; $h_{SP,i}$ between STi and PR; $h_{S,i}$ between STi, $i = 1, ..., K$ and cognitive BS. Moreover, $g_P=|h_P|^2$, $g_{PS,i}=|h_{PS,i}|^2$, $g_{SP,i}=|h_{SP,i}|^2$ and $g_{S,i}=|h_{S,i}|^2$ denote the respective fading power gains.

As leader in the sequential game, the primary link announces its selection of time ratio α and the rule of relay selection, the cognitive BS follows and announces its selection of the number of cognitive relays $N=|\mathbf{S}^C|$, finally cooperative CRs decide their transmission power $\mathbf{p}=(p_1,...,p_N)$, $\mathbf{p}\in[0,P_{\max}]^N$ (the maximal allowed power of each nodes is $P_{\max}$) through power control. Notice that each of the N CRs in $\mathbf{S}^C$ is constrained to use the same power for both cooperation and own traffic (we introduce this constraint lest a selfish CR would contribute little to in earning spectrum opportunities, yet use full power for own traffic. See last two subslots in Fig. 1-(b) and Fig. 1-(c)). Assuming that primary link keeps the total consumed energy in a slot as constant, and the transmission power of the primary when not leasing is expressed as $P_p^0$. Finally, the single-sided spectral density of the independent white Gaussian noise at the receivers (both of primary and secondary network) is $N_0$.

We are interested in two scenarios characterized by different assumptions about the available channel state information (CSI) at different entities and different constraints on the available transmission strategies:

1) Complete and perfect information (Section 3): in this baseline case, the primary link, the cognitive BS and the CRs are assumed to be aware of the instantaneous realizations of all the channel power gains in the system (i.e., $g_P$, $g_{PS,i}$, $g_{SP,i}$ and $g_{S,i}$), and the payoffs of all players under every strategy profile are common knowledge[26].

2) Incomplete information (Section 4): in this scenario, we remove the idealistic assumptions made in the baseline case discussed above. In particular, we assume that the primary link only knows $g_P$, $g_{PS,i}$ and $g_{SP,i}$, while the secondary BS are only aware of all the fading channels within the secondary network, $g_{S,i}$. In other words, neither primary link nor secondary network has priori knowledge of channels inside the other peer.

## 3. Complete information spectrum leasing model

In this section, we describe and analyze spectrum leasing based on space-time coded cooperation within a sequential game framework under the assumption of complete information. We first discuss the primary part (leader) of the model in subsection 3.1, and the secondary network (follower) part which includes secondary BS and CRs in Sec. 3.2 and 3.3. We then provide a backward induction approach for the hierarchical spectrum leasing model in Sec.3.4.

### 3.1. Primary link model

Based on the available instantaneous CSI, the primary link selects the slot allocation parameters $\alpha$ towards the aim of optimizing its SINR in a slot. To start with, in the baseline case of no leasing, the signal-to-noise-ration (SNR) on the direct link between the PT and PR is $\Gamma^0_{PT,PR} = g_P P^0_P / N_0$, this can be seen as the reservation payoff[26], in other words, the payoffs that primary link can ensure without cooperation.

Instead, if the fraction of time leased to the CRs is $1-2\alpha > 0$, then the transmitting power of PT is now $P_P = P^0_P / \alpha$, and the SNR at PR and STi in the first phase is respectively $\Gamma_{PT,PR} = g_P P_P / N_0$ and $\Gamma_{PT,i} = g_{PS,i} P_P / N_0$. In the second phase, the second nodes assigned as cooperative relays forward the frame received in the first phase to PR. We assume the use of a collaborative scheme based on amplify-and-forward (AF) multihop and space-time coding (authors of [25] declare that AF schemes do not provide the necessary flexibility in slot allocation needed in their scenario. We point out that both DF and AF schemes can be used in our framework, yet we choose AF for the sake of comparison). The PR combines the directly received signal at the PR and relayed signals from $S_i$, using techniques such as maximal ratio combining (MRC). In this case the achievable SINR at the PR reads [13, 20]:

$$\Gamma_{PT,ST_i,PR} = \frac{P_P p_i g_{PS,i} g_{SP,i}}{N_0(P_P g_{PS,i} + p_i g_{SP,i} + N_0)} \qquad (1)$$

In which $p_i \in \mathbf{p}$ is the power of relay i. If there are $N = |\mathbf{S}^c|$ relays helping primary link, the totally achievable SINR at PR is: $\Gamma_{AF}(\mathbf{p}, \mathbf{S}^C) = \Gamma_{PT,PR} + \sum_{i=1}^{|\mathbf{S}^C|} \Gamma_{PT,ST_I,PR}$ (2)

In which $\mathbf{p} \in [0, P_{\max}]^N$. To summarize, the utility of primary link is given by

$$U_P(\alpha, \mathbf{S}^C, \mathbf{p}) = \begin{cases} \Gamma^0_{PT,PR}, & \textbf{\textit{other}} \\ 2\alpha \Gamma_{AF}(\mathbf{p}, \mathbf{S}^C), & \alpha \in (0, 1/2) \end{cases}$$

The primary link aims at solving the following optimization problem:

$$\max_{0<\alpha\leq 1/2} U_P(\alpha, \mathbf{S}^C, \mathbf{p}), \; s.t. \mathbf{S}^C \subseteq \mathbf{S}, \mathbf{p} \in [0, P_{\max}]^N \tag{3}$$

*3.2. The heterogeneous opportunistic power control among cognitive nodes*

➢ *The distributed power control model under heterogeneous QoS requirement*

Given the leasing fraction 1-2α and the set of cooperative relays $\mathbf{S}^C$, we investigate the power control problem among CRs (players) based on static non-cooperating game theory. Based on whether a node is selected to relay for the primary link, the set of CRs can be classified into two types: the set of cooperative cognitive radios (CCRs), denoted as $\mathbf{S}^C$ and the set of non-cooperating cognitive radios (NCCRs), denoted as $\mathbf{S}^{NC} = \mathbf{S} \setminus \mathbf{S}^C$. As the CCRs gain communication opportunity for the entire cognitive network by relaying for primary link, these nodes deserve higher QoS (represented in, e.g., SINR) than NCCRs. That is to say, each class has distinct QoS requirements. The cognitive BS compensates CCR k for its cost (of power in relaying) and contributions (in gaining the communication chance) by setting a target SINR $\Gamma_{\mathbf{S},k}^{th}$, which turns out to be function of $\alpha$ (fraction of slot used by the primary link), $|\mathbf{S}|$ (CRs' total) and $|\mathbf{S}^C|$ (number of CCRs). The difference in QoS requirement of CCRs and NCCRs results in distinct utility functions. The utility of CCR k is given by:

$$U_k^C(\alpha, p_k, \mathbf{p}_{-k}) := \rho_k(\alpha, g_{\mathbf{S},k})\sqrt{\Gamma_{\mathbf{S},k} - \Gamma_{\mathbf{S},k}^{th}(\alpha, |\mathbf{S}|, |\mathbf{S}^C|)} + (1-\rho_k(\alpha, g_{\mathbf{S},k}))p_k, \quad \alpha \in (0, 1/2) \tag{4}$$

In which $\Gamma_{\mathbf{S},k}$ is the achieved SINR of CCR k, and the first term denotes how close the real SINR of CCR k is from its target $\Gamma_{\mathbf{S},k}^{th}$, the second term represents cost of k in power (maybe also in compensating the interference it causes to other nodes). We introduce weighting factor $\rho_k(\alpha, g_{\mathbf{S},k})$ to tradeoff the preference of CCR k between reaching its QoS target and decreasing its cost (in power). Assuming that the BS adopts single-user-decoding and takes signals from other nodes as interference in detecting signal of node k, then $\Gamma_{\mathbf{S},k} = g_{\mathbf{S},k} p_k / (N_0 + \sum_{j \neq k}^N g_{\mathbf{S},j} p_j)$ ($\mathbf{p}_{-k}$ denotes the power strategies of nodes other than k).

The utility function of NCCR k accounts for the difference between the achievable SINR and the cost of transmitted energy:

$$U_k^{NC}(p_k, \mathbf{p}_{-k}) := \rho_k(0, g_{\mathbf{S},k})\sqrt{\Gamma_{\mathbf{S},k}} - (1-\rho_k(0, g_{\mathbf{S},k}))p_k, \quad \alpha \in (0, 1/2) \tag{5}$$

Note that given utility functions defined as in (4) and (5), a CCR should try to minimize its utility, while a NCCR should maximize its utility. Thus the power control game in the cognitive network can be expressed as:

$$\begin{cases} CCR: \min_{0 \leq p_k \leq P_{\max}} U_k^C(\alpha, p_k, \mathbf{p}_{-k}), \; \forall k \in \mathbf{S}^C \\ NCCR: \max_{0 \leq p_k \leq P_{\max}} U_k^{NC}(p_k, \mathbf{p}_{-k}), \; \forall k \in \mathbf{S}^{NC} \end{cases} \tag{6}$$

The optimal solution of model (6) is the function of slot fraction $\alpha$, the particular partition $\mathbf{S}^C$ and $\mathbf{S}^{NC}$ of nodes set $\mathbf{S}$, which can be denoted as $\mathbf{p}^*(\alpha, \mathbf{S}^C, \mathbf{S}^{NC}) = (p_1^*(\alpha, \mathbf{S}^C, \mathbf{S}^{NC}), ... p_K^*(\alpha, \mathbf{S}^C, \mathbf{S}^{NC}))$. It is of worth to note that the exact form of SINR target $\Gamma_{\mathbf{S},k}^{th}(\alpha, |\mathbf{S}|, |\mathbf{S}^C|)$ and weighting factor $\rho_k(\alpha, g_{\mathbf{S},k})$ determines the position of $\mathbf{p}^*(\alpha, \mathbf{S}^C, \mathbf{S}^{NC})$ in the feasible power region $[0, P_{\max}]^K$, and thus influences the achievable SINR of each CR as well as payoff of the primary link exploiting cooperation communication. Further analysis and numerical outcome can be found in subsection 5.1.

➢ *The iterative algorithm*

Using best response (BR) approach [19, 26] on model (6), we construct iterative algorithms both for CCRs and NCCRs:

$$\begin{cases} \textbf{CCR: } p_k^{(n+1)} = \Gamma_{\mathbf{S},k}^{th}(\alpha,|\mathbf{S}|,|\mathbf{S}^C|)\dfrac{I_k^{(n)}}{g_{\mathbf{S},k}} + [\dfrac{\rho_k(\alpha,g_{\mathbf{S},k})}{2(1-\rho_k(\alpha,g_{\mathbf{S},k}))}]^2 \dfrac{g_{\mathbf{S},k}}{I_k^{(n)}} =: \Lambda_k^C(\mathbf{p}) \\ \textbf{NCCR: } p_k^{(n+1)} = [\dfrac{\rho_k(0,g_{\mathbf{S},k})}{2(1-\rho_k(0,g_{\mathbf{S},k}))}]^2 \dfrac{g_{\mathbf{S},k}}{I_k^{(n)}} =: \Lambda_k^{NC}(\mathbf{p}) \end{cases} \quad (7)$$

$$I_k^{(n)} = N_0 + \sum_{j \neq k}^{N} g_{S,j} p_j^{(n)} \quad (8)$$

In which $I_k^{(n)}$ represents noise and interference on signals of node k from other nodes in iteration $(n)$, which corresponding to power control strategies of other nodes. Note that the fixed points (if exist) of iteration algorithm (7) thus formulated coincide with Nash Equilibrium (NE) [26] of game (6).

We might as well assume that $\mathbf{S}^C$ consists of the front N nodes of the K CRs. Then iteration function $\mathbf{\Lambda}(\mathbf{p}):[0,P_{\max}]^K \to [0,P_{\max}]^K$ will produce a sequence of power vectors in the form of $\mathbf{\Lambda}(\mathbf{p}) = (\Lambda_1^C(\mathbf{p}),...,\Lambda_N^C(\mathbf{p}),\Lambda_{N+1}^{NC}(\mathbf{p}),...,\Lambda_K^{NC}(\mathbf{p}))$.

➢ *Convergence analysis*

**Theorem 1** For given $\alpha$ and $\mathbf{S}^C, \mathbf{S}^{NC}$, if a fixed point $\mathbf{p}^*(\alpha,\mathbf{S}^C,\mathbf{S}^{NC})$ exists for the iterative function $\mathbf{\Lambda}(\mathbf{p})$ defined as (7), then this fixed pint is unique, and the power vector sequence $\{\mathbf{p}^{(n)}\}$ produced by $\mathbf{\Lambda}(\mathbf{p})$ converges to $\mathbf{p}^*(\alpha,\mathbf{S}^C,\mathbf{S}^{NC})$.

**Proof**: From Theorem 9 and 10 of [5], we know that if $\mathbf{\Lambda}(\mathbf{p})$ meets the property of two-sided scalability and a fixed point exists for it, then this fixed pint is unique and $\mathbf{\Lambda}(\mathbf{p})$ converges. Thus we need only to prove that $\mathbf{\Lambda}(\mathbf{p})$ satisfies the following two-sided scalability: $\forall \mu > 1, \dfrac{\mathbf{p}}{\mu} \leq \mathbf{p}' \leq \mu\mathbf{p} \Rightarrow \dfrac{\mathbf{\Lambda}(\mathbf{p})}{\mu} < \mathbf{\Lambda}(\mathbf{p}') < \mu\mathbf{\Lambda}(\mathbf{p})$ We first simplify (7) to the form of $\begin{cases} \Lambda_k^C(\mathbf{p}) = a_1 I_k(\mathbf{p}) + a_2 \dfrac{1}{I_k(\mathbf{p})} \\ \Lambda_k^{NC}(\mathbf{p}) = a_3 \dfrac{1}{I_k(\mathbf{p})} \end{cases}$, in which $a_1 = \dfrac{\Gamma_{\mathbf{s},k}^{th}(\alpha,|\mathbf{S}|,|\mathbf{S}^C|)}{g_{\mathbf{S},k}}$, $a_2 = [\dfrac{\rho_k(\alpha,g_{S,k})}{2(1-\rho_k(\alpha,g_{S,k}))}]^2 g_{S,k}$, $a_3 = [\dfrac{\rho_k(0,g_{S,k})}{2(1-\rho_k(0,g_{S,k}))}]^2 g_{S,k}$

Applying the definition of $I_k(\mathbf{p})$ in (8), it's easy to show that $\dfrac{\mathbf{p}}{\mu} \leq \mathbf{p}' \leq \mu\mathbf{p} \Rightarrow \dfrac{I_k(\mathbf{p})}{\mu} < I_k(\mathbf{p}') < \mu I_k(\mathbf{p})$

Firstly, for a CCR k, we have $\dfrac{\Lambda_k^C(\mathbf{p})}{\mu} - \Lambda_k^C(\mathbf{p}') = \dfrac{1}{\mu}(a_1 I_k(\mathbf{p}) + a_2 \dfrac{1}{I_k(\mathbf{p})}) - (a_1 I_k(\mathbf{p}') + a_2 \dfrac{1}{I_k(\mathbf{p}')})$

$= a_1(\dfrac{I_k(\mathbf{p})}{\mu} - I_k(\mathbf{p}')) + a_2(\dfrac{1}{\mu I_k(\mathbf{p})} - \dfrac{1}{I_k(\mathbf{p}')}) = a_1(\dfrac{I_k(\mathbf{p})}{\mu} - I_k(\mathbf{p}')) + a_2(\dfrac{I_k(\mathbf{p}') - \mu I_k(\mathbf{p})}{\mu I_k(\mathbf{p}) I_k(\mathbf{p}')})$

As $\dfrac{I_k(\mathbf{p})}{\mu} - I_k(\mathbf{p}') < 0$, $I_k(\mathbf{p}') - \mu I_k(\mathbf{p}) < 0$, and $a_1$、$a_2$、$\mu$、$I_k(\mathbf{p}')$、$I_k(\mathbf{p})$ are all greater than 0, thus $\dfrac{\Lambda_k^C(\mathbf{p})}{\mu} - \Lambda_k^C(\mathbf{p}') < 0$, namely $\dfrac{\Lambda_k^C(\mathbf{p})}{\mu} < \Lambda_k^C(\mathbf{p}')$. In a similar way as shown before we have $\Lambda_k^C(\mathbf{p}') < \mu \Lambda_k^C(\mathbf{p})$.

Secondly, for a NCCR k, $\frac{\Lambda_k^{NC}(\mathbf{p})}{\mu} - \Lambda_k^{NC}(\mathbf{p}') = \frac{1}{\mu}a_3\frac{1}{I_k(\mathbf{p})} - a_3\frac{1}{I_k(\mathbf{p}')} = a_3(\frac{1}{\mu I_k(\mathbf{p})} - \frac{1}{I_k(\mathbf{p}')})$

$= a_3(\frac{I_k(\mathbf{p}') - \mu I_k(\mathbf{p})}{\mu I_k(\mathbf{p})I_k(\mathbf{p}')}) < 0$ i.e., $\frac{\Lambda_k^{NC}(\mathbf{p})}{\mu} < \Lambda_k^{NC}(\mathbf{p}')$. Similarly we have $\Lambda_k^{NC}(\mathbf{p}') < \mu\Lambda_k^{NC}(\mathbf{p})$.

Thus we conclude that $\Lambda(\mathbf{p})$ satisfies two-sided scalability. According conclusions in theorem 9 and 10 of [5], It is now obvious that the theorem holds. ∎

***Theorem 2*** *For* any given strategy of the primary link and the cognitive BS, there exists fixed points for iterative function $\Lambda(\mathbf{p})$ defined in (7).

***Proof***: According to the definition in (7), it is obvious that $\Lambda(\mathbf{p})$ is continuous at all $\mathbf{p} \geq 0$. Also notice that the iteration is constrained by the maximal transmitting power, i.e. the outputs are up-bounded by $P_{max}$, according to theorem 15 in [5]. It is now obvious that the theorem holds. ∎

Combining theorem 1 and 2, we conclude that there exists unique fixed point for iterative function $\Lambda(\mathbf{p})$ and the iterative process converges to it.

It's worth noticing that when all CRs are chosen to be CCRs, the iteration (7) reduces to $p_k^{(n+1)} = \Gamma_{\mathbf{s},k}^{th}\frac{I_k^{(n)}}{g_{\mathbf{s},k}} + [\frac{\rho_k(\alpha, g_{\mathbf{s},k})}{2(1-\rho_k(\alpha, g_{\mathbf{s},k}))}]^2\frac{g_{\mathbf{s},k}}{I_k^{(n)}}$, which fits into two-sided scalability type of power control; and when neither node cooperates for the primary link (assuming that the CRs are still allowed to transmit), (7) reduces to $p_k^{(n+1)} = [\frac{\rho_k(0, g_{\mathbf{s},k})}{2(1-\rho_k(0, g_{\mathbf{s},k}))}]^2\frac{g_{\mathbf{s},k}}{I_k^{(n)}}$, the type II standard power control. In conclusion, our opportunistic power control algorithm generalizes power control algorithm to support heterogeneous QoS requirements.

*3.3. Cognitive base-station model*

We define utility of cognitive BS (representing the cognitive network) as the difference between total achievable SINR and total power cost of all nodes:

$$U_{CR}(\alpha, \mathbf{S}^C, \mathbf{S}^{NC}, \mathbf{p}) = \sum_{k=1}^{|\mathbf{S}^C|+|\mathbf{S}^{NC}|}(\Gamma_{\mathbf{s},k} - \frac{cp_k}{P_{max}}) \quad (9)$$

In which c>0 is a constant. We brief $U_{CR}(\alpha, \mathbf{S}^C, \mathbf{S}^{NC}, \mathbf{p})$ as $U^{CR}(\alpha, \mathbf{S}^C, \mathbf{p})$ in the following. It is obvious that when the primary link chooses no leasing, the cognitive network has zero utility. The BS determines the number of allowed CCRs by the following model ( ƛ balances the utility of primary and secondary network):

$$\max_{0 \leq |\mathbf{S}^C| \leq |\mathbf{S}|} ƛU_{CR}(\alpha, \mathbf{S}^C, \mathbf{p}) + (1-ƛ)U_P(\alpha, \mathbf{S}^C, \mathbf{p}), \text{ s.t. } 0 < ƛ < 1$$

*3.4. Analysis of spectrum leasing sequential game of complete information*

As the actions of primary (determining $\alpha$), cognitive BS(deciding the number of CCRs) and CRs (setting their transmitting power) take place sequentially and they announce their choices, combining the assumption 1) in subsection 2.2, this model pertains to complete and perfection information dynamic game and we provide the backward induction [26] approach for this particular model in the following:

➢ Firstly we should derive the best response functions (BRF) of CRs given $\alpha$ and $\mathbf{S}^C, \mathbf{S}^{NC}$. We have shown in subsection 3.2 that the aforesaid cognitive power control algorithm converges to the unique fixed point $\mathbf{p}^*(\alpha, \mathbf{S}^C, \mathbf{S}^{NC})$ (denoted as $\mathbf{p}^*(\alpha, \mathbf{S}^C)$ hereafter) of the iteration function. We also use this notation to represent BRF of CRs.

➢ We now considering BRF of cognitive BS given $\alpha$. As BS can theoretically solve power control among the CRs, hence can guess out BRF $\mathbf{p}^*(\alpha, \mathbf{S}^C)$ of the CRs, the problem BS has to solve comes down to:

$$\max_{0 \leq |\mathbf{S}^C| \leq |\mathbf{S}|} \lambda U^{CR}(\alpha, \mathbf{S}, \mathbf{p}^*(\alpha, \mathbf{S}^C)) + (1-\lambda) U_P(\alpha, \mathbf{S}^C, \mathbf{p}^*(\alpha, \mathbf{S}^C)) \tag{10}$$

Given the ordering of candidate CRs by the primary, there are $K = |\mathbf{S}|$ possible size of the CCRs set $\mathbf{S}^C$, and the BS must solve both power control(7) and optimization (10) for K times altogether, and picks the maximal utility (if unique) from them, then the responding number of CCRs is just the solution $|\mathbf{S}^C|^*(\alpha, \mathbf{p}^*(\alpha, \mathbf{S}^C))$ to be found.

➢ Finally, we come to the BR of primary link. Similar to analysis in the above step, the primary can dope out BRF of the BS(i.e., $|\mathbf{S}^C|^*(\alpha, \mathbf{p}^*(\alpha, \mathbf{S}^C))$ )and CRs(i.e., $\mathbf{p}^*(\alpha, \mathbf{S}^C)$ ) to all feasible $\alpha$, the problem of primary link boils down to:

$$\max_{0 < \alpha \leq 1/2} U_p(\alpha, \mathbf{p}^*(\alpha, \mathbf{S}^C), |\mathbf{S}^C|^*(\alpha, \mathbf{p}^*(\alpha, \mathbf{S}^C))) \tag{11}$$

If the maximization (11) also has unique solution (denoted as $\alpha^*(\mathbf{p}^*, |\mathbf{S}^C|^*)$ ), we can get the backward induction solution $(\alpha^*(\mathbf{p}^*, |\mathbf{S}^C|^*), \mathbf{p}^*(\alpha^*, |\mathbf{S}^C|^*), |\mathbf{S}^C|^*(\alpha^*, |\mathbf{S}^C|^*))$ of the sequential game, and this solution falls into subgame-perfect NE[26], which is stable in the sense of NE.

There are several difficulties in the above approach. Firstly, it is almost impossible to give explicit expressions of $\mathbf{p}^*(\alpha, \mathbf{S}^C)$ and $\alpha^*$; secondly, to produce the optimal ordering of candidate CCRs and further the optimal solution $\alpha^*$, the primary must calculate $\alpha^{*,m}(\mathbf{S}^{C,m}), 1 \leq m \leq M$ under all potential $\mathbf{S}^{C,m}$, which totally amounts to $M = \sum_{m=1}^{K-1} 2^m$, and dopes out $|\mathbf{S}^C|^*(\alpha, \mathbf{p}^*(\alpha, \mathbf{S}^C))$ and $\mathbf{p}^*(\alpha, \mathbf{S}^C)$ for all $0 < \alpha < 1/2$ and $\mathbf{S}^C \subseteq \mathbf{S}$. Obviously, although backward induction is theoretically a mature method for complete information dynamic games, yet as $\alpha$ and $\mathbf{p}$ are defined in continuous region while ordering of candidate CCRs belongs to the domain of permutation and combination, further more, the strategies of the primary, the cognitive BS and CRs are coupling, thus it is actually very hard to find the strict optimal solution of our hierarchical model. We omit approximate analysis here, as although this version of model builds a base for further analysis, it's somewhat too Utopian. It worths to point out that the analysis in Section 4 is totally applicable for this scenario.

## 4. Incomplete information spectrum leasing model and its solution

The analysis of this section is based on CSI assumption 2) in subsection 2.2. Here the primary has no priori knowledge as to those channels inside the cognitive network, thus is not able to reckon the strategy profiles of the cognitive CRs. So the backward induction approach in subsection 3.4 doesn't apply here. We simplify the cognitive BS model in subsection 3.3 and propose an approximate solution for this incomplete information version of the spectrum leasing model.

*4.1. The simplified model of the cognitive network*

The model and algorithm of power control among the CRs remain the same as subsection 3.2. We reduce the cognitive BS model in subsection 3.3 such that BS determines the number of CCRs using the following formula $N = |\mathbf{S}^C| = [(1-2\alpha)|\mathbf{S}|]^+$, in which $[x]^+$ denotes taking the minimal integer greater than or equal to x.

*4.2. Solution to the primary model*

Assuming the primary knows the strategy of cognitive BS under the simplified model, and then primary model (3) becomes:

$$\max_{\alpha \in (0,1/2)} U_P(\alpha, \mathbf{S}^C, \mathbf{p}) = \max_{\alpha \in (0,1/2)} \begin{cases} \Gamma^0_{PT,PR}, & \text{other} \\ 2\alpha(\Gamma_{PT,PR} + \sum_{i=1}^{|\mathbf{S}^C|}\Gamma_{PT,ST_i,PR}), & \alpha \in (0,1/2) \end{cases} \quad s.t. \mathbf{S}^C \subseteq \mathbf{S}, \mathbf{p} \in [0, P_{\max}]^N \quad (12)$$

Problem (12) can be decomposed into three parts: 1, how to determine relay power $p_i, i=1,...,N$; 2, how to order candidate relays; 3, how to find the suboptimal slot fraction $\tilde{\alpha}^*$. We discuss them separately in the following.

➤ *Determining $p_i, i=1,...,N$*

According to the incomplete information assumption, the primary has no priori knowledge of $g_{\mathbf{S},i}, j=1,...K$ and thus is not able to reckon $\mathbf{p}^*(\alpha, \mathbf{S}^C)$. We consider two alternative schemes: 1, setting $p_i, i=1,...,N$ according to uniform distribution in $[0, P_{\max}]^N$; 2, setting all $p_i, i=1,...,N$ equal.

➤ *Ordering candidate relays*

**Scheme 1**, *Relay-SINR(R-SINR) scheme*: Ordering the candidate relays decreasingly according to $\Gamma_{PT,ST_i,PR}$. We have the following conclusion:

**Theorem 3** Given the candidate relays set $\mathbf{S}$ and slot fraction $\alpha$, ordering $\mathbf{S}$ decreasingly according to $\Gamma_{PT,ST_i,PR}$ is equivalent to the optimal ordering under utility maximization.

*Proof:* To produce the optimal ordering rule under utility maximization, in principle, we should compare the primary utility under all possible $\mathbf{S}^{C,m}$ (which total up to $M = \sum_{m=1}^{K-1} 2^m$) and then extract the optimal ordering of the candidate relays, as pointed out in subsection 3.4. Assuming that $(\mathbf{S}^*) = (S_{1^*}, S_{2^*},...,S_{K^*})$ is a permutation of all candidates relays ordering decreasingly according to $\Gamma_{PT,ST_i,PR}$ and $(\mathbf{S}) = (S_1, S_2,...,S_K)$ is any permutation other than $(\mathbf{S}^*)$. For given $\mathbf{S}$ and $\alpha$, $N=[(1-2\alpha)|\mathbf{S}|]^+$ ($0 \le N \le |\mathbf{S}|$) is the number of CCRs given by BS, $(\mathbf{S}^C)^*$ is the set of front N elements of $(\mathbf{S}^*)$ and $(\mathbf{S}^C)$ is the set of front N elements of $(\mathbf{S})$. It's easy to show that $\sum_{i=1}^{|(\mathbf{S}^C)^*|}\Gamma_{PT,ST_i,PR} \ge \sum_{k=1}^{(|\mathbf{S}^C|)}\Gamma_{PT,ST_k,PR}$, Noticing that relays only show in the second term of (12), thus we also have $U_P(\alpha, (\mathbf{S}^C)^*, \mathbf{p}) \ge U_P(\alpha, (\mathbf{S}^C), \mathbf{p})$. i.e., ordering $\mathbf{S}$ decreasingly according to $\Gamma_{PT,ST_i,PR}$ is equivalent to the optimal ordering under utility maximization∎

According to theorem 3, we know scheme R-SINR is equivalent to the optimal ordering. Further more, this ordering can be finished within $O(|\mathbf{S}|\log(|\mathbf{S}|))$ (using, for instance, QKSORT [27]). When combined with the above two power approximating schemes, we get *Relay-Random-power-SINR* (abbreviated to *RR*) and *Relay-Same-power-SINR* (abbreviated to *RS*) scheme.

**Scheme 2**, *Channel-Product (CP) scheme*: Noticing contribution of each relay to primary utility is limited by both $g_{PS,i}$ and $g_{SP,i}$, an instinct scheme is to ordering $\mathbf{S}$ according to $g_{PS,i}g_{SP,i}$.

➤ *Approximating $\alpha^*$ by segmental border values*

For given number $N^* = [(1-2\alpha)|\mathbf{S}|]^+$ of CCRs by BS and the corresponding CCRs set $(\mathbf{S}^C)^*$, the primary can pick slot fraction within $\frac{1}{2}(1-\frac{N^*-1}{|\mathbf{S}|}) > \alpha \geq \frac{1}{2}(1-\frac{N^*}{|\mathbf{S}|})$. Thus (12) can actually be divided into $|\mathbf{S}|+1$ sub-problems: $\max_{\frac{1}{2}(1-\frac{N^*-1}{|\mathbf{S}|}) > \alpha \geq \frac{1}{2}(1-\frac{N^*}{|\mathbf{S}|})} U_P(\alpha,(\mathbf{S}^C)^*,\mathbf{p})$

$$= \max_{\frac{1}{2}(1-\frac{N^*-1}{|\mathbf{S}|}) > \alpha \geq \frac{1}{2}(1-\frac{N^*}{|\mathbf{S}|})} \begin{cases} \Gamma^0_{PT,PR}, & \text{other} \\ 2\alpha(\frac{g_{PS}P_P}{N_0} + \sum_{i=1}^{N^*} \frac{P_P p_i g_{PS,i} g_{SP,i}}{N_0(P_P g_{PS,i} + p_i g_{SP,i} + N_0)}), \alpha \in (0,1/2) \end{cases} \quad s.t. 0 \leq N^* \leq |\mathbf{S}|, \mathbf{p} \in [0, P_{\max}]^{N^*} \quad (13)$$

And $\alpha^* = \arg\max_{\alpha_i^*} U_P(\alpha_i^*, (\mathbf{S}^C)^*, \mathbf{p})$, in which $\alpha_i^*, 0 \leq i \leq |\mathbf{S}|$ are optimal solutions of the $|\mathbf{S}|+1$ sub-problems as (13).

One simplification for these sub-problems is to simply investigate the border slot fraction $\hat{\alpha} = \frac{1}{2}(1-\frac{N^*}{|\mathbf{S}|})$ and the corresponding border utility for each given $N^*$: $\hat{U}_P(\frac{1}{2}(1-\frac{N^*}{|\mathbf{S}|}), (\mathbf{S}^C)^*, \mathbf{p})$ (14)

Thus we reduce the computing complexity from $|\mathbf{S}|+1$ optimization sub-problems (13) to merely $|\mathbf{S}|+1$ evaluation (14). Pick the biggest value $\hat{U}_P^*$ from these $|\mathbf{S}|+1$ utilities, and the corresponding slot fraction $\hat{\alpha}^*$ is the sub-optimal border approximating.

## 5. Numerical simulation

We considered a 2-km-square cell with cognitive BS centered at the origin and CRs were chosen randomly from a uniform distribution [7] in the square, while PT and PR are located randomly within 1-km-square cell centered at the BS. $P_{\max} = 600mW$. Background receiver noise power within the user's bandwidth of $N_0 = -133dBW$ was used in the simulations. The channel gain was determined according to $h = r^{-a}$, in which r is the distance between the transmitter and the receiver, $a = 2$. We adopt typical value simulation in subsection 5.1 and 5.2 while in 5.3 we use Monte Carlo methods.

*5.1. Simulating the heterogeneous power control in cognitive network*

C. W. Sung etc. point out in [5] that their opportunistic power control algorithm can achieve tenfold throughput increase when compared with target tracking. Yet it magnifies the near–far effect and thus intensifies unfairness. As a generalization of C. W. Sung's algorithm in supporting heterogeneous QoS requirements (of CCRs and NCCRs), our algorithm inherits its merit in throughput while avoids its weakness of unfairness by introducing target SINR $\Gamma^{th}_{\mathbf{S},k}(\alpha,|\mathbf{S}|,|\mathbf{S}^C|)$ and weighting factor $\rho_k(\alpha, g_{\mathbf{S},k})$. We in this subsection simply investigate our iterative algorithm from the view points of converging rate, fairness and the compensation for CCRs etc.

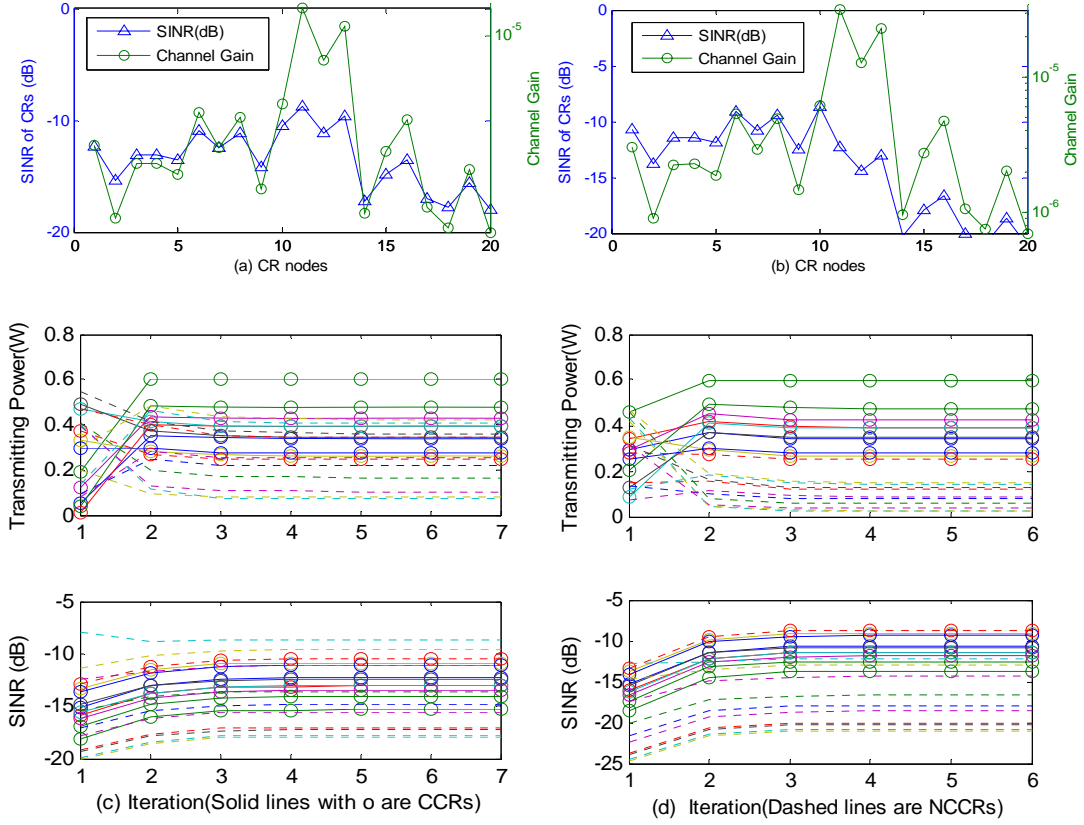

***Fig. 2*** Heterogeneous opportunistic power control (convergence & fairness) *(a)* Received SINR& channel quality of each link pair ($\lambda = 2$) *(b)* Received SINR& channel quality of each link pair ($\lambda = 6$) *(c)* Convergence of power & SINR ($\lambda = 2$) *(d)* Convergence of power & SINR ($\lambda = 6$)

The exact form of SINR target $\Gamma^{th}_{\mathbf{S},k}(\alpha, |\mathbf{S}|, |\mathbf{S}^C|)$ and weighting factor $\rho_k(\alpha, g_{\mathbf{S},k})$ influence position of the fixed point, converging rate and further others. We set $\rho_k(\alpha, g_{\mathbf{S},k}) = \dfrac{2e^{\lambda\alpha/2}}{g_{\mathbf{S},k}^{3/4} + 2e^{\lambda\alpha/2}}$ and $\Gamma^{th}_{\mathbf{S},k}(\alpha, |\mathbf{S}|, |\mathbf{S}^C|) = \dfrac{\lambda\alpha}{\sqrt{|\mathbf{S}|\cdot|\mathbf{S}^C|}}$ (15) in the simulation. Then (4) and (5) come down to respectively: $U^C_k(\alpha, p_k, \mathbf{p}_{-k}) := \dfrac{2e^{\lambda\alpha/2}}{g_{\mathbf{S},k}^{3/4} + 2e^{\lambda\alpha/2}}$ (16)

$$U^{NC}_k(p_k, \mathbf{p}_{-k}) := \dfrac{2}{g_{\mathbf{S},k}^{3/4} + 2}\sqrt{\Gamma_{\mathbf{S},k}} - \dfrac{g_{\mathbf{S},k}^{3/4}}{g_{\mathbf{S},k}^{3/4} + 2} p_k \quad (17)$$

Combining (15) and (7), we get the following iteration:

$$\begin{cases} CCR: p_k^{(n+1)} = \dfrac{\lambda\alpha}{\sqrt{|\mathbf{S}|\cdot|\mathbf{S}^C|}} \dfrac{I_k^{(n)}}{g_{\mathbf{S},k}} + \dfrac{e^{\lambda\alpha}}{g_{\mathbf{S},k}^{3/2}} \dfrac{g_{\mathbf{S},k}}{I_k^{(n)}} := \Lambda_k^C(\mathbf{p}) \\ NCCR: p_k^{(n+1)} = \dfrac{1}{g_{\mathbf{S},k}^{3/2}} \dfrac{g_{\mathbf{S},k}}{I_k^{(n)}} := \Lambda_k^{NC}(\mathbf{p}) \end{cases} \quad (18)$$

In Fig.2 we give typical simulation result where $|\mathbf{S}|=20$ in which 1~10 belong to $\mathbf{S}^c$ and $\alpha=0.25$. 2-(a) and 2-(b) show three properties of our algorithm: 1, the SINRs of all CRs are positively correlated with their channel gains; 2, CCRs always achieve higher SINR than NCCRs under similar channel conditions; 3, We can increase the SINRs of CCRs relative to which of NCCRs by increasing $\lambda$. While 2-(c) and 2-(d) show that our algorithm boasts rapid converging in one hand, and validate the above property 2 and 3 in the other hand.

In Table 1, the Unitary SINR is defined as SINR vs. channel power gain ratio of a node; SINR-power-ratio is just as its name implies; Relative Increment (Relative- INC) means(Value_of_CCR – Value_of_NCCR)/ Value_of_NCCR. With the increasing of $\lambda$, the first column shows that power of CCRs increase while that of NCCRs decrease; the second column shows similar trend of Unitary SINR with the first column; yet the third column shows that the SINR gain of CCR by increasing $\lambda$ is costly in power, i.e., The gain of CCR in SINR accords with law of diminishing marginal utility [26, 28].

*Table 1* Performance comparison between CCRs and NCCRs ($\lambda=2、4、6$ and $8$)

| $\lambda$ | Mean average Transmitting power | | | Mean average Unitary SINR | | | Mean average SINR-power-ratio | | |
|---|---|---|---|---|---|---|---|---|---|
| | CCR (W) | NCCR (W) | Relative-INC | CCR (dB) | NCCR (dB) | Relative-INC | CCR | NCCR | Relative-INC |
| 2 | 0.37549 | 0.24292 | 54.6% | 43.2293 | 41.2472 | 4.81% | 115.126 | 169.794 | -7.6% |
| 4 | 0.37639 | 0.14375 | 95.8% | 44.2075 | 39.8433 | 10.6% | 117.451 | 277.163 | -22.3% |
| 6 | 0.37663 | 0.08513 | 120.0% | 44.892 | 38.1724 | 16.3% | 119.193 | 448.385 | -46.0% |
| 8 | 0.37692 | 0.05077 | 134.3% | 45.3412 | 36.3165 | 21.9% | 120.294 | 715.305 | -83.2% |

From Fig. 2 and Tab. 1, we arrive at the conclusion that our power control algorithm is fast, flexible and fair in that it converges in a few iterations; it can compensate CCRs flexibly and it is capable of overcoming near–far effect.

*5.2. Simulation of the approximating solutions for the primary network*

In this subsection we investigate how close our estimations (using the three schemes given in subsection 4.2) is to the real optimal value (i.e. $\hat{\alpha}^*$ vs. $\alpha^*$ and $\hat{U}_P^*$ vs. $U_P^*$).

The typical simulating results are shown in Fig. 3 in which $|\mathbf{S}|=10$. We see that three estimated primary utility curves do not overlap with each other and they are generally below the respective theoretic utility curves. Numbers inside parentheses of X-axis in (b) denotes the estimated $|\hat{\mathbf{S}}^c|^*$ of each scheme, including *RR*, *RS* and *CP scheme*, which all equal to 4, corresponding to slot fraction $\hat{\alpha}^*=0.3$. Noticing that the three estimated optimal utility $\hat{U}_P^*$ all locate at the summit of each utility curve (proving that the three approximating schemes are indeed valid.), and those of RR and CP are almost same, which are somewhat higher than that of RS. Furthermore, it can be see that the reservation payoff is lower than all utility curves, indicating that cooperative leasing always dominates non-cooperation, no matter how many CRs are chosen as CCRs.

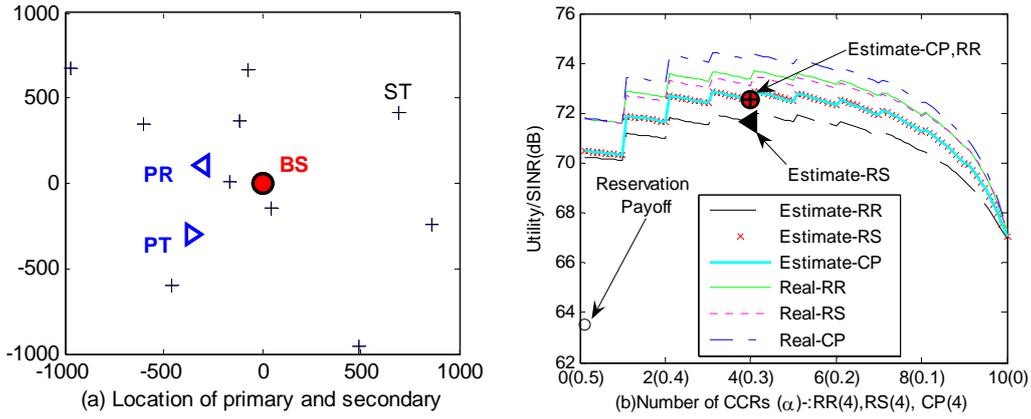

*Fig. 3* Theoretic and estimated primary utility under three estimating schemes (subsection 4.2) vs. number of CCRs (and slot fraction $\alpha$ )

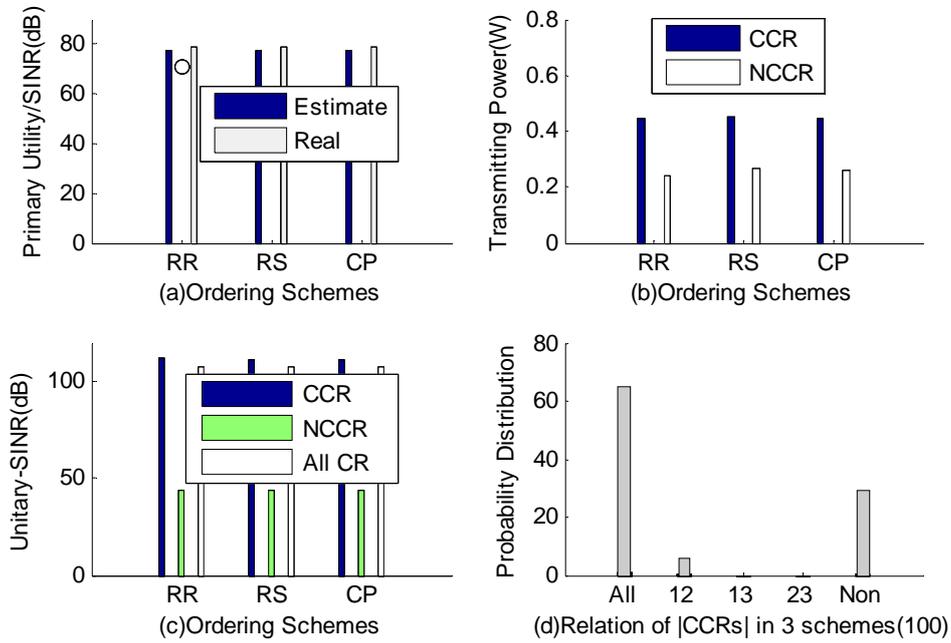

*Fig. 4* Utility of primary link and secondary network (with respect to three approximating schemes, 100 runs) *(a)* The estimated and theoretic utility of primary (o denotes the reservation payoff) *(b)* Transmitting power of CCRs & NCCRs *(c)* Mean average unitary-SINR of CCRs, NCCRs and all CRs *(d)* Do the three approximating schemes produce same $|\widehat{\mathbf{S}}^C|^*$ (1-RR, 2-RS, 3-CP)?

## 5.3. Monte Carlo simulation of primary and secondary cooperative leasing

Each of the above two subsections investigates only the primary link or secondary network. Here we consider performances of both, under our cooperative leasing model. $|\mathbf{S}|=10$, and we repeat our simulation for 100 times. Fig.4 shows that the mean utilities estimated by three approximating schemes are almost equal (4-(a)), all less than the respective theoretic value, and same can be said about power (4-(b)) and Unitary SINR (4-(c)) of CCRs. From 4-(d), we see that within the 100 repeating, there are about 70 times that all three schemes produce equal $|\mathbf{S}^C|$, about 20 times all unequal and about 3 times RR equals RS yet not equal to CP.

To be brief, although the three approximating scheme do not always produce same result, yet it's difficult to distinguish which is better from the view of mean performance.

We conclude this section as follows: 1, both subsection 5.2 and 5.3 indicate that generally our cooperative leasing framework guarantees that cognitive network achieves communication chance while primary link gains utility higher than its reservation payoff. 2, our heterogeneous power control algorithm overcomes the unfairness shortcoming of algorithm in [5] in addition to inherits its high throughput merit; 3, the proposed approximating schemes can achieve valid sub-optimal output and reduce the computational complexity greatly.

## 6. Conclusion

In this manuscript, we propose and analyze a dynamic implementation of hierarchical property-rights model of cognitive radio. We give the backward induction approach for the complete information model and provide approximating algorithm for the incomplete information version. Analysis and numerical results show that our models and algorithms provide a promising framework for fair and effective spectrum sharing.

## Acknowledgments

Supported by National Natural Science Foundation of China (No.60772062), the Key Projects for Science and Technology of MOE (No.206055) and the Key Basic Research Projects for the Natural Science of Jiangsu Colleges (No.06KJA51001).

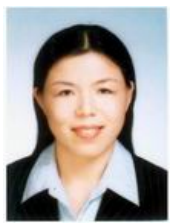
**Li Cuilian**, is currently a lecturer of Zhejiang Wanli University, and Ph.D. candidate of the Institute of Signal Processing and Transmission at Nanjing University of Posts and Telecommunications. She received her master degree in communication engineering from Fudan University in 2004, and her bachelor degree in automation from Harbin Engineering University, China in 1992. Her current research interests focus on cognitive radio and cooperative communication with emphasis on radio resource allocation and management based on game theory. Contact her at licl@njupt.edu.cn

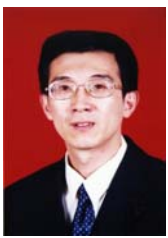
**Yang Zhen**, Ph.D., is a professor and doctoral advisor of Nanjing University of Posts and Telecommunications. His research interests include wireless communications and network signal processing, voice processing and modern voice communication technologies, and information security technologies. Email: yangz@njupt.edu.cn.

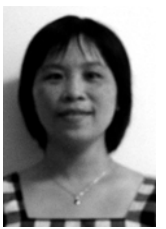
**Li Jun (1971-),** is an associate professor of Zhejiang Wanli University, and PhD candidate of information networking institute at Nanjing University of Posts and Telecommunications. Her research focuses on P2P networks and performance optimization, Internet traffic classification and identification, and distributed network management. Li Jun got her bachelor and master degree in Zhijiang University, China. Contact her at lijunreed@163.com

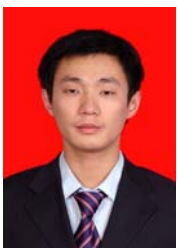
**Tian Fen**, Ph.D., is a teacher at the College of Automation, Nanjing University of Posts and Telecommunications. His research interests include wireless communications and network signal processing, and cognitive radio and radio resource management. Email: tianf@njupt.edu.cn.